# FPA-Debug: Effective Statistical Fault Localization Considering Fault-proneness Analysis

Farid Feyzi, Esmaeel Nikravan, Saeed Parsa
Department of Computer Engineering
Iran University of Science and Technology
Tehran, Iran
farid_feyzi@comp.iust.ac.ir, nikravan@comp.iust.ac.ir, parsa@iust.ac.ir

*Abstract*— **The aim is to identify faulty predicates which have strong effect on program failure. Statistical debugging techniques are amongst best methods for pinpointing defects within the program source code. However, they have some drawbacks. They require a large number of executions to identify faults, they might be adversely affected by coincidental correctness, and they do not take into consideration fault-proneness associated with different parts of the program code while constructing behavioral models. Additionally, they do not consider the simultaneous impact of predicates on program termination status. To deal with mentioned problems, a new 'fault-proneness'-aware approach based on elastic net regression, namely FPA-Debug has been proposed in this paper. FPA-Debug employs a clustering-based strategy to alleviate coincidental correctness in fault localization and finds the smallest effective subset of program predicates known as bug predictors. Moreover, the approach considers fault-proneness of code during statistical modelling through applying different regularization parameter to each program predicates depending on its location within program source code. The experimental results on well-known test suite, Siemens, reveal the effectiveness and accuracy of the FPA-Debug.**

## I. Introduction

Even though software companies make excessive efforts to get rid of software faults during the in-house testing process, they cannot confidently claim that their deployed software is bug free. However, some bugs reveals themselves after the software release during which end users discover and report them to the corresponding software maintenance team [1]. Restrictions on available time, money and people to enhance software on one hand and increases in complexity of software on the other hand, make manual detection and correction of programmatic bugs painstaking and impractical. This has motivated many researchers during the past few years to develop automated software debugging techniques with minimum need for human intervention [2][3][4]. Among fault localization techniques, statistical debugging methods have achieved great success [3][4][12]. In fault localization process, the statistical debuggers evaluate the value associated with program predicates such as branches, loops and function return values in different executions of program and seek to reveal which predicates relate to the program faults [4]. To collect such information, extra code is often inserted before each predicate within the program code. This process is called instrumentation [1]. Considering that bugs are generally nondeterministic, unpredictable and may appear at any time in a program execution, statistical methods can be best applied to analyze the values of the program predicates and construct a model of its behavior. The extracted models can be further applied to discover the program misbehaviors leading to the identification of different types of faults which may appear in the program.

The existing statistical techniques have major limitations. First, a huge number of both failing and passing executions are required in order to perform the statistical analysis. Providing the passing executions is practical, using regression testing prior to the deployment phase of software. But, since the software companies endeavor to eliminate most bugs before the software release, there might be a few number of failing test cases to construct a fault localization model [5]. Constructing statistical model using a few number of failing executions possibly lead to poor support of the model. Thus, we require an approach that could be applicable when the number of predicates of a program is much greater than the number of passing and failing executions.

The second main problem is the coincidental correctness problem [9]. A test is said to be coincidentally correct if it executes the faulty statements but reveals no failure. Statistical fault localization techniques leverage the coverage information to discover the suspicious elements of a program. However, these techniques can be negatively affected by coincidental correctness, when the program produces the correct output while the defect is executed. In fact, when coincidentally correct tests are present, the faulty entity will probably be ranked as less suspicious than when they are not present. The behavior of an erroneous program is unpredictable and uncertain. Therefore, when modeling the run time behavior of a program suspicious to error, uncertainty and randomness in the program state should be considered.

The third issue is that existing statistical approaches are merely based on program execution data and to the best of our knowledge, no method has gained benefit from knowledge on fault-proneness associated with different portions of the code. Our programming experiences has shown that certain areas of the program code might be more likely to be faulty than other areas. For example, a programmer is more likely to perpetrate faults while writing a recursive function than a simple function. Generally speaking, code complexity correlates with the defect rate and robustness of the application [18][19]. Code which is too complex is often the reason for bad code quality and erroneous programs. Complex code is not only error prone, it is also difficult to test. Static code metrics are direct measurements of source code that can be used in an attempt to quantify various software properties. These are properties that may potentially relate to code quality, and therefore to fault-proneness. The fourth important problem is the huge number of predicates which is common in large scale programs. In each run of a program, most of the predicates are logically redundant having no predictive power [1].

To deal with the first and fourth problems, a new statistical approach based on elastic net [7] namely FPA-Debug have been proposed in this paper. The privilege of elastic net is its ability to penalize coefficients in order to minimize the residual sum of squares. The elastic net handles multicollinearity in the variable selection problem by combining the LASSO with ridge regression [13]. Elastic net method has grouping effect which means that it could select the high collinear features and put them into a group. This characteristic is very useful in fault localization techniques, since it could select fault relevant predicates and put them into groups which we call bug predictor group. This is also advantageous when program has several faults. Elastic net could also produce interpretable models and tends to assign high coefficients to most relevant predicates [7]. Put it differently, it ignores redundant and irrelevant predicates. The most important advantage of elastic net is in cases that the number of features is much more than the number of observations.

Before applying the elastic net method, we use a clustering-based strategy to deal with the coincidental correctness problem [9]. The intuition behind the strategy is that tests in the same cluster have similar behaviors. Thus a passed test in a cluster with many failed tests is highly susceptible to be coincidentally correct because it has the potential to execute the faulty elements like failing tests of that cluster.

As we mentioned earlier, certain areas of the program code might be more likely to be faulty than other areas. This fact needs to be considered during the design of fault localization strategies in a way that fault-prone areas of the code be examined with more emphasis while scoring the predicates. To this end, for more fault prone statements we use shrinkage parameter with smaller value comparing to less fault prone statements. As a result, the statements in more fault prone parts of the program are hardly eliminated from the corresponding model considering that in case of to be faulty they may achieve larger fault suspiciousness scores. In contrast, the statements in less fault prone parts of code should fight with each other to obtain large fault suspiciousness scores. Note that this does not mean we neglect the role of statements in less fault prone locations, on program failure. Instead, we emphasize on more caution and carefulness to statements on more fault prone parts of the program.

After applying elastic-net, fault suspicious predicates are selected according to their elastic-net coefficient. Finally, faulty sub-paths including highly fault suspicious predicates are reported as the context of failure. The remaining parts of the paper are organized as follows. In section two, the previous works which have been done in this context are presented. In section three, we describe our proposed approach. The empirical results are shown in section four. Finally, concluding remarks are described in section five.

## II. RELATED WORK

The idea of using a regression method for fault localization was first introduced in [4]. The technique uses a naïve logistic regression to select suspicious features from program runtime profiles and clusters the selected features to identify similar features. Liblit's method in [4], uses a regularized logistic regression to select suspicious predicates which are correlated with software crash. Considering two categories of train and test data set, it maximizes the log likelihood of the training set to learn a classifier which has good predictions on the test data set. As mentioned in [12], Liblit's method contains some deficiencies. For large programs with a huge number of predicates, the majority of predicates are redundant and irrelevant to the failure. Therefore, the regression model in [4] may retain some irrelevant predicates in the model. The main problem with the logistic regression in [4] is that it is not appropriate for programs with a high amount of correlation among predicates. Furthermore, it is incapable to find multiple bugs. Since the traditional regression method is inadequate for applications with highly correlated variables, in [13], a ridge regression model has been applied and [14] introduces a combination of ridge and lasso regression methods. The ridge regression is helpful when there is high correlation among predicates. However, for large programs with huge numbers of predicates, it could not provide interpretable models. The lasso method has the feature selection capability that removes irrelevant features and preserves the significant ones. However, when there is high correlation among predicates, it assigns relatively high coefficient to a single predicate and very small coefficient to the other correlated ones. Furthermore, it does not have grouping effect. Both ridge and lasso methods fail to work for P >> N problems where the number of predicates is much more than the number of executions. This may limit their scalability for large programs and small input data (i.e., test cases). To address these problems, Hierarchy Debug [15] was introduced.

There are other fault localization methods which are not based on regression models. Context aware technique [3] considers the vertical dependence among predicates. It combines the methods of feature selection, clustering, and static control flow graph analysis to identify the failure context. Cause transition algorithm [16] uses the Delta Debugging algorithm to narrow down the state difference between failing and passing runs according to their memory graphs. The strategy is based on determining whether a change in a program state makes a difference in the test outcome.

## III. THE PROPOSED APPROACH

This section presents the main idea of our proposed approach. FPA-Debug has three main stages. These stages are detailed in this section.

### A. Alleviate Coincidental Correctness Problem

Coincidental correctness occurs when a test case executes the faulty elements but no failure is triggered. Previous studies have demonstrated that coincidental correctness is prevalent in both two forms: strong and weak [9], [17]. When coincidentally correct test cases are present, the faulty elements will likely be ranked as less suspicious than when they are not present. As shown in the previous studies [11], [17], the efficiency and accuracy of coverage based fault localization can be improved by cleaning the coincidentally correct test cases. However, it is difficult to identify coincidental correctness because we do not know the locations of faulty elements in advance. In this paper, we leverage a clustering-based strategy to identify the subset of test suite that is possible to be coincidentally correct [9]. We use cluster analysis to group test cases into different clusters. Passed test cases which are grouped into the same cluster with the failed ones are very likely to be coincidentally correct, and are added to the set of identified coincidentally correct tests. The reasons are two-fold: 1) A test case which executes the faulty statement does not necessarily induce a failure, but not vice versa. It is a sufficient condition for a failed test case to execute the faulty statements. 2) It is assumed that test cases with similar execution profiles will be clustered together. Therefore, the identified passed test cases will have similar execution profiles with the failed ones. After identifying the set coincidentally correct tests, relabeling strategy is used in order to improve the effectiveness of elastic net method. To this aim, these tests are relabeled from "Passed" to "failed".

### B. Predicting Fault-prone Areas of the Code

The estimation of a module's fault-proneness is important for minimizing cost and improving the effectiveness of the software testing process. Software complexity metrics are often used as indirect metrics of reliability since they can be obtained relatively early in the software development life cycle. Using complexity metrics to identify components which likely contain faults allows software engineers to focus the verification effort on them, thus achieving a reliable product at a lower cost. In this paper, we assume that there exists a relationship between the measures of software complexity and the faults found during testing and operation phases. The foundation of this assumption comes from past research which has given empirical evidence of the existence of this relationship.

Static code metrics are direct measurements of source code that can be used in an attempt to quantify various software properties. These are properties that may potentially relate to code quality, and therefore to defect-proneness. Perhaps the most well-known static code metrics are based on lines of code (LOC) counts, and give an indication of software size. While LOC-count-based measures aim to provide insight into software size, another set of metrics, proposed by Maurice Halstead in 1977, also aim to provide insight into code complexity and developer effort [18]. The length metric is the sum of all operators and operands, and is an alternate size measure to those based around LOC-counts. The vocabulary metric is the sum of all unique operators and operands; code with a high vocabulary is thought to be hard to read and therefore difficult to maintain. The volume metric describes information content in bits, and is another size related measure. The difficulty metric was claimed to measure how difficult the code was to write, and therefore how error-prone it is likely to be. The complement of this is the level metric, with a lower level thought to indicate less error-prone code. The effort metric is used to measure the effort to comprehend and therefore maintain code, while the content metric was claimed to be a language independent complexity measure. Perhaps the most unjustified of these measures are the error estimate and time to program ones, as both include unfounded constants. Metrics concerned solely with code complexity were proposed by Thomas Mc-Cabe in 1976 [19]. The most well used of these is known as the cyclomatic complexity, and is based on program control flow. This metric measures linearly independent paths, and is equal to the upper bound of required unit tests for basis path coverage. When analyzing static code metrics, it is important to know the level of granularity at which they were captured. Common granularities include the file and package level, as well as the module level. In this paper the term module is used as a generic term to refer to the function or method level. Because static code metrics are calculated through the parsing of source code, their collection can be automated. Thus it is computationally and resourcefully feasible to calculate the metrics of entire software systems, irrespective of their size. The NASA Metrics Data Program (MDP) Repository[1] contains module-level data sets explicitly intended for software metrics research. Each data set represents a NASA software system/subsystem and contains the static code metrics and fault data for each comprising module.

We used a logistic regression to construct the fault prediction model from static code metrics. Most linear modeling applications, such as regression analysis, can produce unstable models when the independent variables have a strong relationship between themselves. In this cases, principal component analysis can be used to reduce the dimensions of the metric space and obtain a smaller number of orthogonal domain metrics. In our experiment, the principal component analysis applied on the 11 complexity metrics revealed three distinct complexity domains, having eigenvalues greater than 0.9. The regression model was built based on the domain metrics which have been generated from the principal component analysis.

---

[1] http://mdp.ivv.nasa.gov.

*C. Statistical Fault-localization Method*

To analyze the simultaneous impact of predicates on each other and on the program termination status, a logistic regression method can be applied to model the runtime behavior of a program in failing and passing runs. In this paper we have proposed a method based on elastic net [7], a well-known shrinkage method to select most effective bug predictors. The privilege of shrinkage methods is their ability to penalize coefficients in order to minimize the residual sum of squares [8]. The elastic net method provides a stable estimation method that may be used to advantage when there is high multicollinearity among predictor variables [9]. The elastic net estimators are stable in the sense that they are not affected by slight variations in the estimation data [1]. Elastic net method has grouping effect which means that it could select the high collinear features and put them into a group. This characteristic is very useful in fault localization techniques, since it could select fault relevant predicates and put them into groups which we call bug predictor group. This is also advantageous when program has several faults. Elastic net could also produce interpretable models and tends to assign high coefficients to most relevant predicates [1]. In other words, it ignores redundant and irrelevant predicates. The most important advantage of elastic net is in cases that the number of features is much more than the number of observations.

In order to understand the role of elastic net method in the proposed fault localization technique, we first describe the problem. In order to find a relationship between program predicates and the failing or passing state of the program, a linear regression model could be constructed. Let $P_i = (p_{i1}, p_{i2}, ... p_{in})$ be the vector of predicates' values and $y_i$ be the corresponding program termination status for the $i$th execution. The relationship between $P_1, P_2, ..., P_m$ (m is the number of program executions) and the program termination status Y is formulated as

$$Y = \beta_0 + \beta_1 P_1 + \beta_2 P_2 + ... + \beta_n P_n + \varepsilon \quad (1)$$

The term $\beta_0$ in equation (1) is the intercept, also known as the bias in machine learning. $\beta_i$'s are constants known as regression coefficients (weights) and $\varepsilon$ is the error of model. The elastic net is a shrinkage method which could be effective when the number of features is more than the number of observations. It also has grouping effect which gives groups of high correlated features. These privileges make the elastic net a very good method for selecting bug predictors in the program.

The $\beta$ parameters in elastic net method are estimated as

$$\hat{\beta} = \arg\min_\beta \sum_{i=1}^{m}(y_i - \beta_0 - \sum_{j=1}^{n} p_{ij}\beta_j)^2 \quad (2)$$

Subject to $\lambda \sum_{j=1}^{n} \beta_j^2 + (1-\lambda)\sum_{j=1}^{n} \beta_j \leq r$

$r$ is a tuning parameter which depends on the dataset features. As shown in equation (1), elastic net, imposes two penalties on the coefficients. The parameter $\lambda$ in penalty function is in [0,1) in most applications. If $\lambda = 0$, the equation (1) becomes the lasso equation and if $\lambda = 1$, it would be the ridge regression [14]. Thus, by choosing an appropriate value of $\lambda$ between zero and one, the equations may have the characteristics of both lasso and ridge regression methods [14]. There are different well-known methods to select such tuning parameters [8]. In the case we have only training data, cross-validation method is an appropriate one to estimate the prediction error and compare different models. For the two tuning parameters in the elastic net, we applied cross-validation on a two-dimensional surface. For choosing r, we examine different values of $\lambda$, such as 0, 0.001, 0.05, 0.1, …, 0.95 and for each one we tried 5 and 10 fold cross validation dependent on the training data. The tuning parameter, r, is the one giving the smallest miss-classification error. An important characteristic of elastic net is its grouping power which finds group of highly correlated features. Consider a program has multiple faults and each fault is manifested in one or more predicates.

*D. Considering Fault proneness*

Glmnet package [20], allows us to apply separate penalty factors to each coefficient. Its default is 1 for each parameter, but other values can be specified. In particular, any covariate with penalty factor equal to zero is not penalized at all. Let $v_j$ denote the penalty factor for $j$th variable. The penalty term becomes

$$\lambda \sum_{j=1}^{p} v_j P_\alpha(\beta_j) = \lambda \sum_{j=1}^{p} v_j [(1-\alpha)\frac{1}{2}\beta_j^2 + \alpha |\beta_j|] \quad (3)$$

This is very useful when we have prior knowledge or preference over the variables. In many cases, some variables may be so important that one wants to keep them all the time, which can be achieved by setting corresponding penalty factors to 0. Having obtained fault-proneness likelihood of predicates using static metrics, we apply equation (3) in order to define separate penalty factor for each covariate in elastic net regression model.

$Penalty.factor_i =$

$1 - (FP(s_i))^{3 \leq \theta \leq 4}$, if $FP(s_i) \leq 0.5$ (4)

$1 - (FP(s_i) * 0.2 \leq \theta \leq 0.3)$, if $FP(s_i) > 0.5$

## IV. EMPIRICAL RESULTS

To show the effectiveness of proposed approach, we conducted an experiment on the Siemens programs. The Siemens set contains seven C programs, and all of them can be downloaded from the SIR repository [21]. Each of the programs has a correct version, a number of faulty versions seeded with a single fault, and a corresponding test suite. A brief description of Siemens test suites is presented in Table I.

A well-known evaluation framework is the T-score measure which has been previously applied in [15][16] With this framework, the static dependencies of a program to be debugged are explored to compute the distance between a faulty statement and the statement(s) reported by a fault localization algorithm. In this framework, for a faulty program Q, the corresponding program dependence graph, PDG, is defined with the following properties: 1) each node of the graph corresponds to a particular program statement. 2) An edge between two nodes represents the control or data dependence between their corresponding statements. For statements which are reported as fault suspicious by a fault localization algorithm, the corresponding nodes can be labeled as suspicious nodes. In a similar way, the node(s) corresponding to the actual faulty statements are labeled failure origin node(s). A human debugger starts from the suspicious node(s) and while traversing the graph in breadth first order, the number of nodes that should be searched to reach one of the failure origin nodes is computed. The set of statements covered by BFS is called $N_{examined}$.

TABLE I. A BRIEF DESCRIPTION OF SIEMENS TEST SUITES

| Program | Description | #test cases | #Lines of Code | # faulty versions |
|---|---|---|---|---|
| print tokens | Lexical analyzer | 4056 | 472 | 7 |
| print tokens2 | Lexical analyzer | 4071 | 399 | 10 |
| replace | Pattern replacement | 5542 | 512 | 32 |
| schedule | Priority scheduler | 2650 | 292 | 9 |
| schedule2 | Priority scheduler | 2680 | 301 | 10 |
| tcas | Altitude separation | 1578 | 141 | 11 |
| tot info | Information measure | 1054 | 440 | 41 |

Assuming that PDG has N nodes, the T-score is computed as follows:

$$T - Score = \left(\frac{N_{examined}}{N}\right) * 100 \quad (5)$$

A. *Comparison with fault localization algorithms*

In this part, we compare the performance of FPL-Debug with different fault localization algorithms, using a number of existing evaluation frameworks. The aim is to evaluate the performance of the proposed technique from different aspects. First, we compare our approach and five well-known fault localization techniques according to the T-score measure. The results of this comparison are presented in Figure1. As shown in Figure1, the performance of FPL-Debug in finding faults with regard to T-score values outperforms other algorithms. FPL-Debug is applied on 106 out of 132 versions of the Siemens suite. It has detected 50 faults with less than 1 % manual code examination. Therefore, we may conclude that a success key of FPL-Debug is considering the simultaneous effect of predicates on each other and on the program termination state as well as taking into account fault-proneness during construction of statistical model. With the same T-score value, Tarantula detected 17 faults considering that it instruments all statements in a given program and hence it imposes relatively high overhead on program execution. With a T-score less than 10 %, FPL-Debug detected 98 faults dramatically outperforming other algorithms.

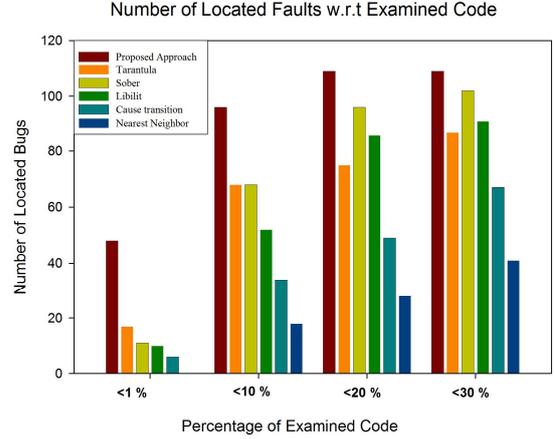

Fig. 1. A detailed performance comparison between methods according to T-Score framework

Another evaluation framework is P-score which has been introduced in [22]. It measures the number of examined predicates instead of examining statements. The P-score value for a single fault case is computed as follows:

$$P - score = \frac{\Pr ed - index\ in\ L}{|L|} * 100\% \quad (6)$$

Where Pred-index, starting from one, is the order of the actual fault relevant predicate in L, the list of the ranked predicates given by the fault localization algorithm. For example, if the first predicate in the list of the ranked predicates is the actual fault relevant predicate, Pred-index would be one. The actual fault relevant predicate is the predicate that is nearest to a faulty code in a given program. The smaller the value of the P-score, the more the fault localization algorithm will be effective.

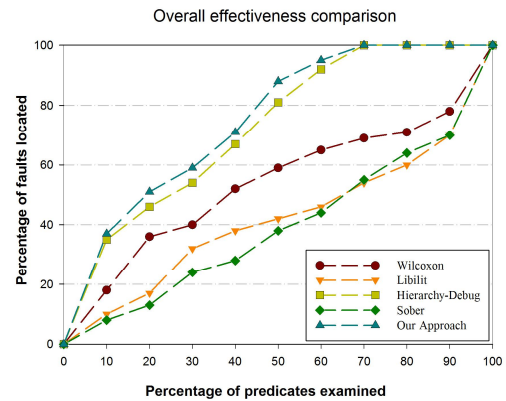

Fig. 2. A detailed performance comparison between methods according to P-Score framework

Figure 2 presents the overall results of FPL-Debug and the four mentioned algorithms, according to the P-score framework, on all programs of the Siemens suite. As shown in the figure, in 38 faulty versions of the Siemens suite, FPL-Debug locates the faulty predicate with less than 10% predicate examination.

V. CUNCLUDING REMARKS

In this paper a new combinatorial approach to fault localization is proposed. The combinatorial approach considers both dynamic and static attributes of the subject program to rank predicates according to their effects on the program termination states. Fault-proneness of different portions of the code is considered in statistical modeling and the experimental results conclude that the proposed approach is able to localize more faults compared to previous techniques with less amount of code inspection by the programmer.

REFERENCES

[1] B. Liblit. "Cooperative bug isolation", PhD thesis, University of California, Berkeley, Springer, 2004.
[2] H. Cheng, D. Lo, Y. Zhou, X. Wang. "Identifying Bug Signatures Using discriminative Graph Mining," International Symptoms on Software esting and Analysis, pp. 141-151, ACM Press, 2009.
[3] L. Jiang, Z. Su. "Context-aware statistical debugging: From bug predictors to faulty control flow paths" In Proceedings of 22st IEEE/ACM international conference on automated software engineering, pp. 184–193, 2007.
[4] B. Liblit, A. Aiken, X. Zheng, M.I. Jordan. "Bug isolation via remote program sampling", In Proceedings of the ACM SIGPLAN conference on programming language design and implementation, pp. 141–154, 2003.
[5] A. Zeller. "Why Programs Fail: A Guide to Systematic Debugging", Morgan Kaufmann, San Francisco, 2006.
[6] X. Wang, S. C. Cheung, W. K. Chan and Z. Zhang. "Taming coincidental correctness: Coverage refinement with context patterns to improve fault localization", in Proc. 31st Int. Conf. on Software Engineering, pp. 45-55, 2009.
[7] C. Del Mol, E. De Vito, L.Rosasco. "Elastic-Net regularization in Learning Theory", technical report, MIT press, 2008.
[8] S. Chatterjee, A. Hadi, B. Price. "Regression Analysis by Example", 4th edn, Wiley Series in Probability and Statistics, New York, 2006.
[9] Y. Miao et al. "A clustering-based strategy to identify coincidental correctness in fault localization." International Journal of Software Engineering and Knowledge Engineering, pp. 721-741, 2013.
[10] Dickinson, W., Leon, D., & Podgurski, A. "Finding failures by cluster analysis of execution profiles", Proceedings of the 23rd international conference on Software engineering. IEEE Computer Society, 2001.
[11] Podgurski, et al. "Automated support for classifying software failure reports", Software Engineering, Proceedings. 25th International Conference on. IEEE, 2003.
[12] B. Liblit, M. Naik, A. Zheng, A. Aiken, and M. Jordan. Scalable statistical bug isolation. In Proc. of ACM SIGPLAN 2005 Int. Conf. on Programming Language Design and Implementation (PLDI'05), 2005.
[13] S. Parsa, M. Vahidi-Asl, S. Arabi Naree. "Finding causes of software failure using ridge regression and association rule generation methods." Software Engineering, Artificial Intelligence, Networking, and Parallel/Distributed Computing, SNPD'08. Ninth ACIS International Conference on. IEEE, 2008.
[14] S. Parsa, et al. "Statistical software debugging: From bug predictors to the main causes of failure." Applications of Digital Information and Web Technologies, ICADIWT'09. Second International Conference on the. IEEE, 2009.
[15] S. Parsa, M. Vahidi-Asl, M. Asadi-Aghbolaghi. "Hierarchy-Debug: a scalable statistical technique for fault localization." Software Quality Journal 22(3), pp.427-466, 2013.
[16] Cleve, Holger, and Andreas Zeller. "Locating causes of program failures."Proceedings of the 27th international conference on Software engineering. ACM, 2005.
[17] W. Masri and R. A. Assi, "Cleansing test suites from coincidental correctness to enhance fault-localization", in Proc. 3rd Int. Conf. on Software Testing, Verifcation and Validation, pp. 165-174, 2010.
[18] Halstead, Maurice H. "Elements of Software Science". (Operating and programming systems series). Elsevier Science Inc., 1977.
[19] McCabe, T.J., "A complexity measure. IEEE Transactions on Software Engineering", SE-2 (4), 308–320, 1976.
[20] J. Friedman, T. Hastie, R. Tibshirani. "Regularization Paths for Generalized Linear Models via Coordinate Descent", Journal of Statistical Software, 33(1), pp.1-22, 2010.
[21] G. Rothermel, S. Elbaum, A. Kinneer, and H. Do, "Software-artifact infrastructure repository", http://sir.unl.edu/portal ,2006.
[22] [22] Z Zhang, Z., Chan, W. K., Tse, T. H., Hu, P., & Wang, X. "Is non-parametric hypothesis testing model robust for statistical fault localization?" Journal of Information and Software Technology, 51, 1573–1585, 2009.